\documentstyle[preprint,epsfig,aps]{revtex}
\begin{document}
\title{An Analytical Approach to Fluctuations in Showers}

\author{R.A. V\'azquez}

\address{INFN sez. di Roma and Universit\'a della Basilicata}

\maketitle

\setcounter{page}{0}
\thispagestyle{empty}

\begin{abstract}
We review the problem of fluctuations in particle shower theory. By using a
generalization of Furry equation, we find relations between the $n$--particle
correlation function and the number of particles average or 1--particle
correlation function. Such relations show that the average is the only
independent dynamical variable. We also develop a numerical code to solve the
equation for the correlation functions and compare the results with those from
a Montecarlo simulation which show a perfect agreement between both methods.

\end{abstract} \newpage

\section{Introduction}

The problem of fluctuations in cascade showers has been the subject of 
extensive analytical studies in the past. The work on analytical calculation
was pioneered by Furry \cite{Furry} followed by work by Kolmogorov {\it et al.}
\cite{Kolmogorov} and Janossy \cite{Janossy}. In the early 60's, Kalmykov
\cite{Kalmykov} calculated the second moments for hadronic processes. The books
of Ramakrishnan \cite{Rama} and Harris \cite{Harris}, and the review of Messel
\cite{Messel}, and others (see references in \cite{Messel}) are a good review
of the subject. More recently it was considered in \cite{Lagutin,Uchaikin}.

Nowadays, this analytical approach may look only an academic problem, 
considering the complexity achieved by existing MonteCarlo programs, where
every single detail is included. It is difficult to think of how to improve
these methods. However one should not forget that MonteCarlo simulations are
not free of drawbacks. For instance, at extremely high energies, of the order
of the Greisen-Zatsepin cut-off, the number of particles is so large that
treating the whole shower in the MonteCarlo scheme is not possible. The usual
way-out is to implement some kind of ``thinning'' procedure where part of the
particles below a certain thinning threshold are dropped, and the rest is kept
with an increased weight \cite{Hillas}. By construction, this procedure yields
the correct average number of particles, but largely overestimates the
fluctuations. Increasing the statistics, these fake fluctuations should
disappear, making the problem not so serious. More serious would be spectrum
deformations in the fluctuation of particles, which should be checked in an
independent way. Also systematics in the determination of shower parameters
\cite{Parente}, specially the energy, could be induced by this procedure. Other
methods which are combination of MonteCarlo and analytical calculations are
possible \cite{Zatsepin}.

It should be clear that a method to estimate fluctuations independently of
MonteCarlo codes is not only convenient but also necessary. In this paper we
present a general overview of the subject together with a new formulation based
on the use of functional probabilities, which yields the equations for the
moments. For the sake of simplicity the problem is treated for the case of only
one type of particle and only one possible type of interaction. More realistic
cases are left for future work \cite{fluc_next}. To make clearer the
exposition, we will start with a extremely simple case. We will consider the
case where the energy of every particle is neglected, we will call this case
the ``zero-dimensional'' case. This case can be solved completely and will give
some insight into the full problem, where the energy is taken into account.

The functional technique, although more complicated in principle, has the
advantage of leading to a linear equation for the probability, generalizing the
Furry equation, and making it possible to get a general formal solution.
Although we will not make any further study of these possibilities, we think it
is worthwhile to notice this point. 

This work is divided in three sections. In section \ref{trivia} we consider the
simple case of zero dimensions, i.e. the case where only total number of
particles is counted, disregarding their energy. Although it has little
interest by itself we will treat this problem carefully, at length. At the end
of the section we will understand clearly the problem of fluctuations in
showers. This study permits to advance the formulation, and some new surprising
results, for the more realistic case of ``one dimension'', where the energy is
considered. This later case is treated in section \ref{1d}. In section
\ref{numerical} we give some numerical results and we finish with some
conclusions.

\section{A trivial example} \label{trivia}

We will start with the simple case of the Furry distribution. It represents a
physical system where particles can split in two with probability $w$ by unit
length. Therefore the Furry distribution corresponds to a cascade shower in
zero dimensions. If we choose the unit length such that $w=1$, the Furry
equation for the probability function $\Pi$ of having $n$ particles at a
distance $t$ reads:
\begin{equation}
\frac{d \Pi(n,t)}{dt} = - n \Pi(n,t) + (n-1) \Pi(n-1,t).
\label{furry0}
\end{equation}
The physical meaning of this equation is straightforward. If we choose $dt$
very small, there are only two possible terms in the change of probability: the
transition $n \rightarrow n+1$ or the transition $ n-1 \rightarrow n$. Each
term in the above equation represents the corresponding transition probability.
The coefficients $n$ and $n-1$ are just the correct combinatorial factors. It
is easy to show that eq. (\ref{furry0}) conserves probability. Defining:
\begin{equation}
f_0= \sum_{n=0}^\infty \Pi(n,t),
\label{ptot}
\end{equation}
we can see, by using eq. (\ref{furry0}), that $f_0$ verifies:
\begin{equation}
\frac{d f_0}{dt} = \sum_0^\infty \left[ -n \Pi(n,t) + (n-1) \Pi(n-1,t)
\right] = - \langle n\rangle + \langle n\rangle = 0.
\label{furry_conservation}
\end{equation}
For the initial condition of one particle at zero depth, 
$\Pi(n,t=0)=\delta(n-1)$, Furry equation has the solution:
\begin{equation}
\Pi(n,t)=e^{-t} (1-e^{-t})^{n-1},
\label{furry_solution}
\end{equation}
as can be checked directly. The average number of particles grows
exponentially, without limit, representing the multiplicative character of a
cascade process.

In the early 50's Janossy \cite{Janossy} showed that there exist an alternative
formalism to treat this problem. His method is usually called the
``regeneration point method'' and it is, somewhat, alternative to Furry's one. 
Let's consider the initial particle at $t=0$ and consider the first interaction
point, let's say, at $t=\zeta$, where the particle splits into two. Then the
probability of having $n$ particles at $t$ is given by the probability that
each of the two new branches produce a shower with $n'$ and $n-n'$
respectively, suming over all possible values of $n'$ and integrating over all
possible values of $\zeta$. This gives the following equation for $\Pi$:
\begin{equation}
\Pi(n,t)=\int_0^t d \zeta e^{-(t-\zeta)} \sum_{n'=1}^{n-1} \Pi(n',\zeta)
\Pi(n-n',\zeta).
\label{janossy0_preview}
\end{equation}
Multiplying by $\exp(t)$ and differentiating with respect to $t$ we obtain, 
after some algebra
\begin{equation}
\frac{d \Pi(n,t)}{dt} = - \Pi(n,t) + \sum_{n'=1}^{n-1} \Pi(n',t) \Pi(n-n',t).
\label{janossy0}
\end{equation}
Which is the Janossy equation for this case. We can check directly that the one
particle solution eq. (\ref{furry_solution}) is also a solution to the Janossy
equation, as it should. 

The first thing to notice is that eq. (\ref{janossy0}) is not linear in $\Pi$.
This implies that probability is {\it not} automatically conserved. Indeed we
find:
\begin{equation}
\begin{array}{ll}
{\displaystyle 
\frac{d f_0}{dt}} & {\displaystyle = - \sum_{n=0}^\infty \Pi(n,t) + 
\sum_{n=0}^\infty \sum_{n'=1}^{n-1} \Pi(n',t) \Pi(n-n',t) } \\
& {\displaystyle
 = -f_0+ \sum_{n_1=0}^\infty \sum_{n_2=0}^\infty \Pi(n_1,t) \Pi(n_2,t) } \\
& {\displaystyle = -f_0 + f_0^2, }
\end{array}
\label{janossy_conservation}
\end{equation}
{\it i.e.} the total probability is conserved only in the case $f_0=1$. This
should be compared to Furry's case. In that case the conservation of the total
probability was automatic, independently of the initial value of this quantity.
One could say that global normalization was arbitrary. In Janossy's case in
order to represent a shower, we need to fix the normalization. This is a
consequence of the non-linear structure of eq. (\ref{janossy0}). It may have
consequences in the one-dimensional case (see below).

In the case of one dimension, we will hardly be able to calculate the full
probability $\Pi(n,t)$, and we will content ourselves with the calculation of
the first few moments of the distributions. 

Let's define the $k$-particle correlation function as 
\begin{equation}
f_k(t)= \langle n^k(t) \rangle = \sum_0^\infty n^k \Pi(n,t).
\end{equation}
And let's evaluate the equations for the first moment or mean value, and the
second moment or mean squared value, separately for Furry and Janossy's cases
in some detail. In Furry's case the mean value verifies :
\begin{equation}
\begin{array}{ll}
{\displaystyle
\frac{d f_1(t)}{dt}} &{\displaystyle = \sum_0^\infty n \left( - n \Pi(n,t) + (n-1) \Pi(n-1,t) 
\right) } \\
 & {\displaystyle 
= - \sum_0^\infty n^2 \Pi(n,t) + \sum_0^\infty n (n-1) \Pi(n-1,t) } \\
& {\displaystyle 
= - \sum_0^\infty n^2 \Pi(n,t) + \sum_0^\infty (n-1)^2 \Pi(n-1,t) + 
\sum_0^\infty (n-1) \Pi(n-1,t) = f_1(t) }
\end{array}
\label{average0}
\end{equation}
while for the mean squared value 
\begin{equation}
\begin{array}{ll}
{\displaystyle
\frac{d f_2(t)}{dt}} & {\displaystyle = \sum_0^\infty n^2 \left( - n \Pi(n,t) + (n-1) \Pi(n-1,t)
\right) } \\
& {\displaystyle
= - \sum_0^\infty n^3 \Pi(n,t) + \sum_0^\infty n^2 (n-1) \Pi(n-1,t) } \\
& {\displaystyle
=- \sum_0^\infty n^3 \Pi(n,t) + \sum_0^\infty (n-1)^3 \Pi(n-1,t) } \\
& \;\; {\displaystyle
+ \sum_0^\infty ( 2 (n-1)^2 + (n-1) ) \Pi(n-1,t) = 2 f_2(t) + f_1(t). }
\end{array}
\label{square0_furry}
\end{equation}
For Janossy's approach we find after some similar algebra that the equation for
the average value is the same as eq. (\ref{average0}), however for the mean
square number we find:
\begin{equation}
\frac{d f_2(t)}{dt} = f_2(t) + 2 f_1(t)^2.
\label{square0_janossy}
\end{equation}
Which is different from eq. (\ref{square0_furry}). All other higher orders
equations are also different for the two cases. In fact it is not difficult to
evaluate them. We find for the moment of order $k$ for Furry's case:
\begin{equation}
\frac{d f_k(t)}{dt}= \sum_{j=0}^k 
\left(\begin{array}{c} k \\[-0.5cm] j \end{array}\right) f_{j+1}(t),
\end{equation}
while for the Janossy's:
\begin{equation}
\frac{d f_k(t)}{dt}= -f_k(t) + \sum_{j=0}^k 
\left(\begin{array}{c} k \\[-0.5cm] j \end{array}\right)
f_{k-j}(t)
f_j(t).
\end{equation}
Although at first it may look surprising, a little thought should clarify the 
problem. Eqs. (\ref{furry0}) and (\ref{janossy0}) are {\it different}, so that
although the solution for one particle in the initial state is the same, they
can not have all solutions equal for arbitrary number of particles in the
initial state. Alternatively one can say that the Janossy's formalism describes
a cascade shower only in the case of one particle in the initial state. Other
solutions of the Janossy equation are not physical showers. 

This being true, both equations should give the same answer for the case of one
particle in the initial state. In other words, for such initial condition the
equations for the moments should be compatible. Such compatibility condition,
imposes severe restrictions on the possible form of the solutions. For
instance for the moments of order 2, it implies that:
\begin{equation}
2 f_2(t)+ f_1(t)=f_2(t)+ 2 f_1(t)^2,
\end{equation}
or
\begin{equation}
f_2(t)=2 f_1(t)^2 - f_1(t).
\label{solution0}
\end{equation}
One can check directly that the solution (\ref{solution0}) is an actual
solution of both the equations (\ref{square0_furry}) and
(\ref{square0_janossy}), if $f_1(t)$ satisfy eq. (\ref{average0}).
Compatibility of both equations give the actual solution of higher moments,
without need to solve the differential equation. This property is in fact true
for all the higher moments, {i.e.} any moment, $f_k$, can be written as a
polynomial of order $k$ in $f_1(t)$:
\begin{equation}
f_k(t)=\sum_{s=1}^k C(k,s) f_1(t)^s
\label{solutionk}
\end{equation}
The proof for this case is very simple and we won't give it here, $C(k,s)$ are
a set of (constant) coefficients. The important point is that in the
one--dimensional case a similar equation must also hold, as we will see, making
the evaluation of moments easier. We do not need to solve a different
differential equation for each moment. We must point out that the in solution 
(\ref{solutionk}) we are not using the Green's function properties of the
system. Note that eq. (\ref{solutionk}) is {\it local} in time.

The solution to the system of equations for $f_k(t)$ given by eq.
(\ref{solutionk}) may look less surprising if we consider the complete solution
given by eq. (\ref{furry_solution}). Indeed we see that eq. 
(\ref{furry_solution}) can be expressed as:
\begin{equation}
\Pi(n,t)=e^{-t} (1-e^{-t})^{n-1}= \frac{1}{f_1(t)} (1- \frac{1}{f_1(t)})^{n-1},
\end{equation}
that is to say, the probability distribution $\Pi(n,t)$ has only one free
parameter, $f_1(t)$, and therefore any other moment can be written as a
function of this one parameter. Of course, the fact that this function is a
polynomial of order $k$ for the moment of order $k$ is a lucky property of the
system. In the case of one dimension this property is by no means evident, and
in fact, we were unable to prove it for a momentum of arbitrary order but we
can guess it after this convincing argument. Also, as we will see, it is 
verified for the second order momentum.

\section{Fluctuations in Showers} \label{1d}

We want to generalize eq. (\ref{furry0}) to the case of a continuous dependence
of the number of particles with the energy, and describe a real shower. We will
consider only one type of particle and only one type of process: one particle
splits into two dividing its energy with a given probability distribution. 
Although simple, this model has already all the complications associated with
cascade processes. 

As an intermediate step we consider the case of discretized bins of energy.
Let's assume that we have a system with $s$ bins of energy, so that the state
is defined by the number of particles in each bin, $n_i, i=1,\ldots,s$. The
transition probability by unit of time of one particle in state $i$ to give two
particles in states $j$ and $k$ is given by $w_{ijk}$. Choosing as before a
$dt$ sufficiently small we can only have two possible transitions. And we get
the equation for the probability: 
\begin{equation}
\frac{dP({n_\alpha},t)}{dt}=-\sum_{i,j,k} w_{ijk} n_i P({n_\alpha},t) +
\sum_{i,j,k} w_{ijk} m_i P({m_\alpha},t),
\end{equation}
where $m_\alpha$, given by:
\begin{equation}
m_\alpha=n_\alpha+\delta(\alpha-i)-\delta(\alpha-j)-\delta(\alpha-k),
\end{equation}
is the equivalent of the ``$n-1$'' factor which appears in the eq.
(\ref{furry0}). 

It is straightforward to generalize this equation to the case of a continuous
energy. Let us define the probability of interaction by unit depth by:
\begin{equation}
w(E_0,E_1,E_2)=\frac{d \sigma}{d E_1 d E_2}.
\end{equation}
Where $E_1$ and $E_2$ are the energy carried by the secondaries. Obviously
$E_1+E_2=E_0$, but we will leave the delta function of conservation of energy
implicit in the definition of $w$. Also let's define:
\begin{equation}
w(E_0,E_1)=\int_0^\infty dE_2 \:w(E_0,E_1,E_2),
\end{equation}
and finally:
\begin{equation}
w(E_0)=\int_0^\infty dE_1 dE_2 \:w(E_0,E_1,E_2)=1. 
\end{equation}
Let us call $N(E) dE$ the (random) number of particles with energy between $E$
and $E+dE$. The equation for the probability of having the function $N(E)$ at
depth $t$ is then given by:
\begin{equation}
\begin{array}{ll}
{\displaystyle
\frac{d {\cal P}(N(E),t)}{d t} =} & 
{\displaystyle \int_0^\infty dE_0 dE_1 dE_2 \: \left[
- N(E_0)\: w(E_0,E_1,E_2)\: {\cal P}(N(E),t) \right. } \\
& {\displaystyle
\left. + M(E_0)\: w(E_0,E_1,E_2)\: {\cal P}(M(E),t) \right], }
\end{array}
\label{furry1}
\end{equation}
Where $M(E)$, as before, is given by:
\begin{equation}
M(E)=N(E)+ \delta(E-E_0) - \delta(E-E_1) - \delta(E-E_2).
\end{equation}
${\cal P}(N,t)$ must be understood as a functional defined over the space of
functions $N(E)$. For a given function $N(E)$, ${\cal P}(N(E),t)$ gives the
probability that such a function gives the actual number of particles at depth
$t$ in a given shower. The above equation indeed represents a generalization of
the Furry equation. This equation can be fully justified by taking the
continuum limit of the discretized case when $s \rightarrow \infty$. 

Solving the equation (\ref{furry1}) completely is a very difficult task, 
although it can be used to generate the equations for the moments of the
distribution. We define:
\begin{equation}
f_1(z,t)=\langle N(z,t)\rangle = \int {\cal D}N(E) N(z) {\cal P}(N(E),t),
\label{average_def}
\end{equation}
as the mean value of the number of particles. Then multiplying eq.
(\ref{furry1}) by $N(z)$ and integrating over all $N$ we get for the left hand
side:
\begin{equation}
\int {\cal D}N(E)\: N(z)\: \frac{{\cal P}(N(E),t)}{d t} = \frac{d f_1(z,t)}{d t}
\end{equation}
and the right hand side gives:
\begin{equation}
\begin{array}{rr}
{\displaystyle
\int {\cal D}N(E) N(z)} & 
{\displaystyle \int_0^\infty dE_0 dE_1 dE_2 \left[ - N(E_0)\:
w(E_0,E_1,E_2)\: {\cal P}(N(E),t) \right. } \\
& {\displaystyle
\left. + M(E_0)\: w(E_0,E_1,E_2)\: {\cal P}(M(E),t) \right] } \\
{\displaystyle
= \int_0^\infty dE_0 dE_1 dE_2} & {\displaystyle \left[ - \int {\cal
D}N(E)\: N(z)\: N(E_0)\: w(E_0,E_1,E_2) \:{\cal P}(N(E),t) \right. } \\
& {\displaystyle
+ \int \left. {\cal D}N(E) \:N(z) \:M(E_0)\: w(E_0,E_1,E_2)\:
 {\cal P}(M(E),t) \right] },
\end{array}
\end{equation}
substituting $N(z) = M(z)-\delta(z-E_0)+\delta(z-E_1)+\delta(z-E_2)$ in the
second term in the right hand side, we get for this term:
\begin{equation}
\begin{array}{ll}
{\displaystyle
\int_0^\infty dE_0 dE_1 dE_2} & {\displaystyle \left\{
- \int {\cal D}N(E)\: N(z)\: N(E_0) \:w(E_0,E_1,E_2)\: 
{\cal P}(N(E),t) \right. } \\
& {\displaystyle
\left. + \int {\cal D}M(E) \: M(z) \: M(E_0)\: 
w(E_0,E_1,E_2) \: {\cal P}(M(E),t) \: \right\} } \\
{\displaystyle
+ \int_0^\infty dE_0 dE_1 dE_2} &\: {\displaystyle w(E_0,E_1,E_2) \:
 \int {\cal D}N(E)\: 
[-\delta(z-E_0) + \delta(z-E_1) + \delta(z-E_2) ] 
M(E_0)\: {\cal P}(M,t). }
\end{array}
\end{equation}
The first term of this equation is identically zero, the second term gives the 
following contribution, after some algebra:
\begin{equation}
- f_1(z,t) + 2 \int_0^\infty dE \: w(E,z) \: f_1(E,t).
\end{equation}
So, finally, the equation for the mean number of particles reads:
\begin{equation}
\frac{d f_1(z,t)}{d t}=- f_1(z,t) + 2 \int_0^\infty dE \: w(E,z) \:f_1(E,t).
\end{equation}
Which are the well-known shower equations for this system. We can proceed
further and calculate the equations for the higher moments. Let us define:
\begin{equation}
G(z_1,z_2,t)=\langle N(z_1,t)N(z_2,t)\rangle = \int {\cal D}N(E) \: N(z_1) 
\: N(z_2) \: {\cal P}(N(E),t)
\end{equation}
Then proceeding in the same way and after some algebra we find:
\begin{equation}
\begin{array}{ll}
{\displaystyle
\frac{d G(z_1,z_2,t)}{dt} = } 
& {\displaystyle -2 G(z_1,z_2,t) + 2 \int_0^\infty dE \: w(E,z_1)
\: G(E,z_2) + 2 \int_0^\infty dE \: w(E,z_2) \: G(z_1,E) } \\
& {\displaystyle 
+ \delta(z_1-z_2) \left[
f_1(z_1,t) + \int_0^\infty dE \: w(z_1,E)\: f_1(E,t) \right] } \\
& {\displaystyle - w(z_1,z_2)\: f_1(z_1,t)
- w(z_2,z_1)\: f_1(z_2,t) + w(z_1+z_2,z_1,z_2)\: f_1(z_1+z_2,t) }
\end{array}
\label{correlation}
\end{equation}
Here we see that apart from the evolution term, it appears a source term given
by an operator acting on the mean number of particles. Higher moments will have
more complicated source terms which depended on all the moments of inferior
order. Given the eq. (\ref{correlation}) then the variance of the number of
particles above a certain energy $E_t$ is given by:
\begin{equation}
\langle N(E_t)^2\rangle = \int_{E_t}^\infty d z_1 \int_{E_t}^\infty d z_2 
G(z_1,z_2,t).
\end{equation}
Evolution equations for the higher order moments could be obtained by the same
procedure. 

In general solving eq. (\ref{correlation}) is difficult. Eq.
(\ref{correlation}) gets simplified a little by defining the factorial moment,
$f_2(z_1,z_2,t)$, as:
\begin{equation}
G(z_1,z_2,t)=f_2(z_1,z_2,t)+ \delta(z_1-z_2) f_1(z_1,t)
\end{equation}
and we get:
\begin{equation}
\begin{array}{ll}
{\displaystyle
\frac{d f_2(z_1,z_2,t)}{dt} =} & {\displaystyle -2 f_2(z_1,z_2,t) + 
2 \int_0^\infty dE \: w(E,z_1) \: f_2(E,z_2) + 
2 \int_0^\infty dE \: w(E,z_2) \: f_2(z_1,E) } \\
& {\displaystyle + w(z_1+z_2,z_1,z_2) \: f_1(z_1+z_2,t). }
\end{array}
\label{factorial}
\end{equation}
This type of equations were discussed by Ramakrishnan \cite{Rama} and also by 
Messel \cite{Messel}.

If $w(E,z_1,z_2)$ is homogeneous, {\it i.e.} it only depends on the ratios
$x=z_1/E,y=z_2/E$ then the conventional method to solve this type of equations
is to make a Mellin transformation on the two energy variables. Defining:
\begin{equation}
Z(s_1,s_2)=\int_0^1 dx dy \: w(x,y) \: x^{s_1-1} \: y^{s_2-1},
\end{equation}
and also
\begin{equation}
Z(s)=\int_0^1 dx \: w(x) \: x^{s-1},
\end{equation}
we find the equation for the Mellin transform of $f_2$:
\begin{equation}
\frac{d \bar f_2(s_1,s_2,t)}{dt}=-2 \bar f_2(s_1,s_2,t) + 2 [Z(s_1)+Z(s_2)]\: 
\bar f_2(s_1,s_2,t) + Z(s_1,s_2)\: \bar f_1(s_1+s_2-1,t).
\end{equation}
Which can be integrated easily.
\begin{equation}
\bar f_1(s,t)= E_0^{s-1} e^{-t[1-2 Z(s)]},
\end{equation}
\begin{equation}
\begin{array} {ll}
{\displaystyle \bar f_2(s_1,s_2,t)=} &
{\displaystyle E_0^{s_1+s_2-2} \frac{2 Z(s_1,s_2)}{2 Z(s_1) + 2 Z(s_2)- 
2 Z(s_1+s_2-1) -1} } \\
& 
{\displaystyle 
\left\{e^{-t[2-2 Z(s_1)-2 Z(s_2)]}-e^{-t[1-2 Z(s_1+s_2-1)]} \right\} }.
\end{array}
\end{equation}
for a shower initiated by a single particle of energy $E_0$. However, doing the
Mellin transform back is a difficult task, the usual saddle point 
approximation does not give correct results \cite{Messel}. Besides, it is
difficult to generalize for non-homogenous cross sections. In the case of
electromagnetic showers this would correspond to the A approximation, which is
known to have large corrections. For numerical results it is better to solve
equation (\ref{factorial}) directly, without going through the Mellin
technique.

Instead of doing the full numerical calculation, let us go back to the Janossy
formalism. This case is simpler, because it does not use the functional 
integration. Let us call $\phi(E,n,t)$ the probability of having $n$ particles
with energy above $E$ at depth $t$ in a given shower. The generalization of
equation (\ref{janossy0}) is given by \cite{Janossy}:
\begin{equation}
\begin{array}{ll}
{\displaystyle \frac{d \phi(E,n,t)}{dt}=} &
{\displaystyle -\phi(E,n,t)} \\
&
{\displaystyle +\sum_{n'=1}^{n-1} \int_0^\infty
\int_0^\infty dE_1 dE_2 \: w(E,E_1,E_2)\: \phi(E_1,n',t) \: \phi(E_2,n-n',t) }.
\end{array}
\end{equation}
Again multiplying by $n$ and summing over all $n$ we get the equation for the
mean value of the integrated number of particles. Which gives:
\begin{equation}
\frac{d n(E,t)}{dt}= -n(E,t)+ \int_0^1 dx\: x \: w(x)\: n(E/x,t),
\end{equation}
and multiplying by $n^2$ we get the equation for the mean square number, $n_2$:
\begin{equation}
\begin{array} {ll} 
{\displaystyle
\frac{d n_2(E,t)}{dt}=} & 
{\displaystyle
-n_2(E,t)+ 2 \int_0^1 dx \: x \: w(x) \: n_2(E/x,t) } \\ 
& {\displaystyle +
2 \int_0^1 dx \: x \: w(x)\: n(E/x,t)\: n(E/(1-x),t)}.
\end{array}
\label{square1_janossy}
\end{equation}
As was expected it is different from eq. (\ref{correlation}) and quadratic in 
the mean number like eq. (\ref{square0_janossy}). Notice also that the above 
equations are for the integrated number of particles. 

Equation (\ref{square1_janossy}) only depends on one energy, instead of the
case of Furry, which depends on two. So that one may think that this equation
is best suitable for using the Mellin technique. This is not the case, even in
the ideal case of homogeneous functions, for the source term is not
homogeneous. For numerical evaluations, however, it is easier to treat this
equation. 

In section \ref{trivia} we constructed a solution to the differential equations
without need to integrate them. Similarly, the general solution to the 
equation for the correlation, (\ref{factorial}), can be written as :
\begin{equation}
\begin{array} {ll} 
{\displaystyle 
f_2(E_1,E_2,t)= } & 
{\displaystyle \int \frac{dx}{x} \int \frac{dy}{y} \: \alpha(x,y)
\: f_1(x E_1,t) \: f_1(y E_2,t)} \\ 
& {\displaystyle +
\int \frac{dE'}{E'^2}\: \beta(E'/E_1,E'/E_2)\: f_1(E',t)} .
\end{array}
\label{solution1}
\end{equation}
Where $\alpha(x,y)$ is an arbitrary function which is determined by the 
initial conditions while $\beta(x,y)$ is independent of the initial conditions 
and depends only on the dynamics of the problem. Introducing the general 
solution eq.(\ref{solution1}) in eq.(\ref{factorial}) we obtain the following
constraint on $\beta$, 
\begin{equation}
\beta\left(\frac{1}{x},\frac{1}{y}\right)= -2 \delta(1-x-y) \: w(x) +\int
d\zeta \:w(\zeta)\: \left[ \frac{1}{\zeta} 
\beta\left(\frac{\zeta}{x},\frac{\zeta}{y}\right) -
\beta\left(\frac{\zeta}{x},\frac{1}{y}\right) - 
\beta\left(\frac{1}{x},\frac{\zeta}{y}\right) \right]. 
\label{alpha}
\end{equation}

In the case of one particle of fixed energy as initial condition, 
$\alpha(x,y)$ is given by $\alpha(x,y)=-\beta(x,y)$ and therefore obtaining the
correlation function is equivalent to solving eq.(\ref{alpha}). This solution
can also be obtained from the condition of equivalence between Furry's and
Janossy's equations as we did for eq.(\ref{solution0}) in the zero-dimension
case. 

This solution, in contrast to the attempt of solving directly 
eq.(\ref{solution1}), presents the advantage of being more or less independent
of the homogeneity of the cross-sections. It seems sensible to presume that
similar equations should hold for the higher order moments. 

In principle one can try to solve eq. (\ref{alpha}) perturbatively, the first
order being
\begin{equation}
\alpha_0\left(\frac{1}{x},\frac{1}{y}\right)=-\beta_0\left(\frac{1}{x},\frac{1}{y}\right)
=2 \delta(1-x-y) w(x),
\end{equation}
which would yield a solution (\ref{solution1}) easy to calculate. In practice 
we find that this is not a realistic approximation, giving an incorrect 
behaviour and overestimating the size of fluctuations. 

\begin{figure}
\epsfxsize=10cm
\begin{center}
\mbox{\epsfig{file=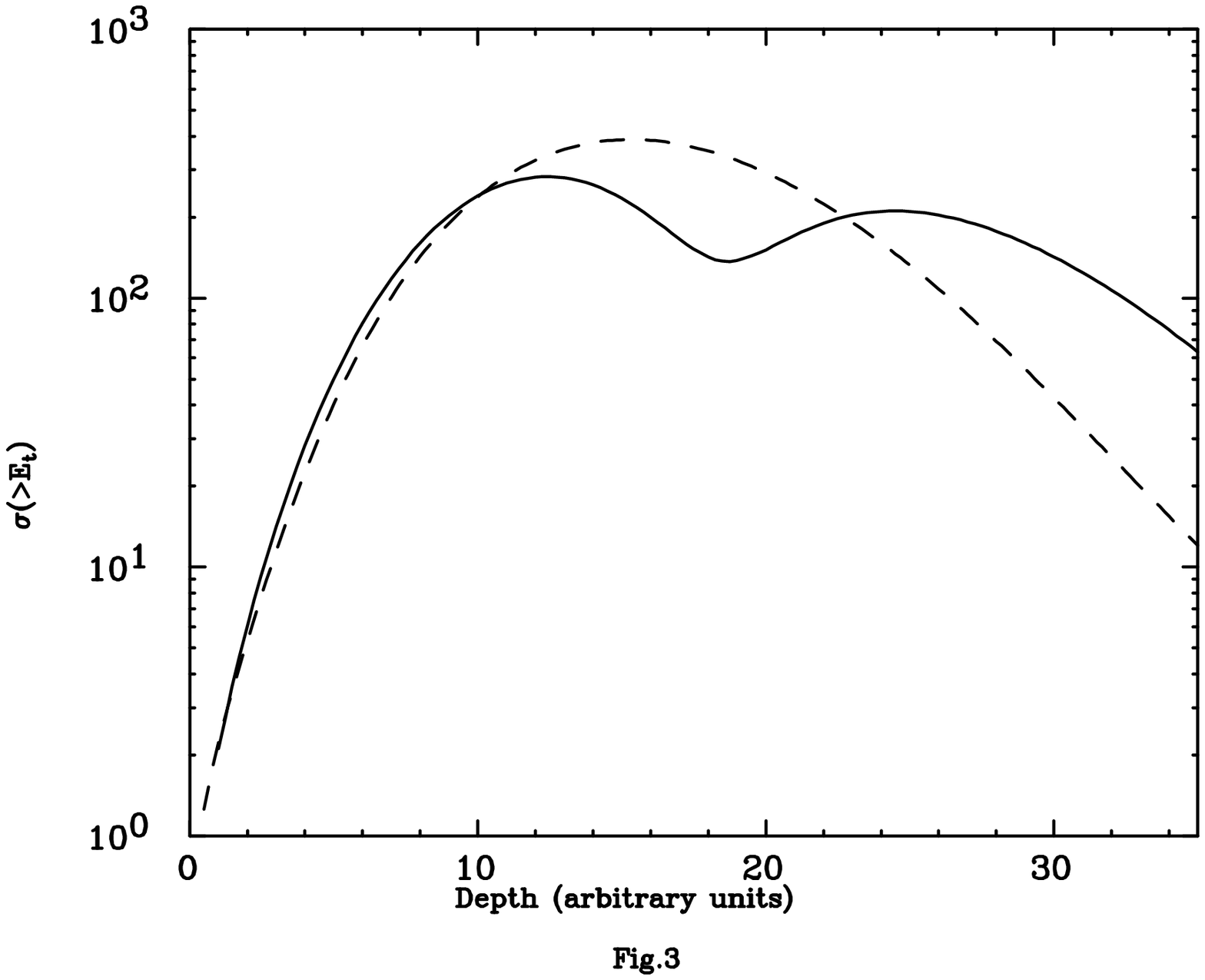,bbllx=60pt,bblly=200pt,bburx=600pt,%
bbury=600pt}}
\end{center}
\caption{
$\sigma$ as a function of depth evaluated by using the Runge-Kutta evolution
code (continuous line) and the approximation, eq. (45) (dashed) for a shower of
initial energy $E=10^4 E_t$. 
}
\label{tres}
\end{figure}

We found that a more reasonable approximation is:
\begin{equation}
\tilde\alpha\left(\frac{1}{x},\frac{1}{y}\right)=
-\tilde \beta\left(\frac{1}{x},\frac{1}{y}\right)=\delta(1-x-y) w(x),
\end{equation}
which gives a good approximation for small depths. Our results show that at the
depth where the average number of particles reaches its maximum value, this
approximation overestimates the size of the fluctuations for a factor of 1.5
--2. At larger depths, it fails completely, underestimating the fluctuations.
In figure (\ref{tres}) we show the above approximation for $\sigma(t)$ and
compare it with the exact result. So that the above equation could be taken
only as a ``rough and ready'' formula to estimate the size of the fluctuations
at early stages of the shower development. Its simplicity is nevertheless
remarkable. 

\section{Results} \label{numerical}

\begin{figure}
\epsfxsize=10cm
\begin{center}
\mbox{\epsfig{file=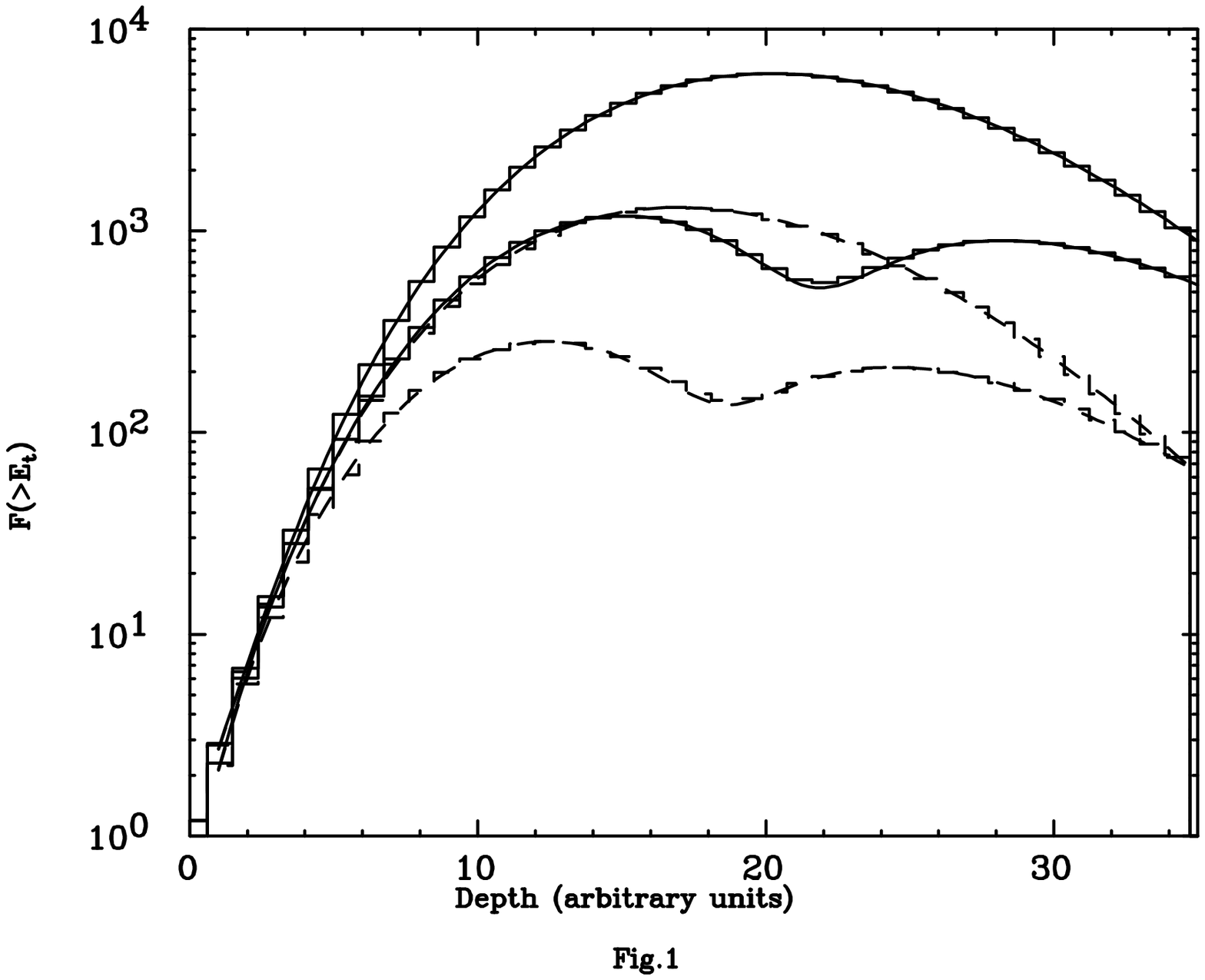,bbllx=60pt,bblly=200pt,bburx=600pt,%
bbury=600pt}}
\end{center}
\caption{
Average number of particles and $\sigma$ as a function of the depth.
Histograms are the results of the MonteCarlo simulation (with a statistics of 
$10^4$ showers). Lines are the result of the evolution code, shown for two 
different energies, $E=10^5 E_t, 10^4 E_t$ where $E_t$ is the threshold energy.
Curves for $\sigma$ are those with two maximum.
}
\label{uno}
\end{figure}

We are now ready to study quantitatively the results of our analysis. In order
to do so, we have developed a code to solve eq. (\ref{square1_janossy}) 
numerically and we have compared the results with those of a MonteCarlo
generator. The program is based in a Runge--Kutta method with interpolation in
the energy. It calculates the mean number of particles and the mean square
number for the simple model given above for arbitrary cross section $w(x)$. We
find that the results agree extremely well with the MonteCarlo, at the level of
per mil accuracy. The results are shown in the figures. In figure (\ref{uno})
we show the average number of particles above energy $E_t$ as a function of the
depth. Also we show the standard deviation, $\sigma$ defined, as usual:
\begin{equation}
\sigma(t)^2=\langle N^2(E_t,t)\rangle -\langle N(E_t,t)\rangle ^2.
\end{equation}
Histogram lines show the MC result whereas continuous lines show the numerical
calculation results of the equation. We see that the agreement is good. 
$\sigma(t)$ shows a double maximum and a minimum near the shower maximum. This
is a general characteristic of shower development, independent of the model and
the cross sections. The reason is that the maximum of $\langle
N^2(E_t,t)\rangle $ is reached at a different depth than the maximum of the
mean number.

\begin{figure}
\epsfxsize=10cm
\begin{center}
\mbox{\epsfig{file=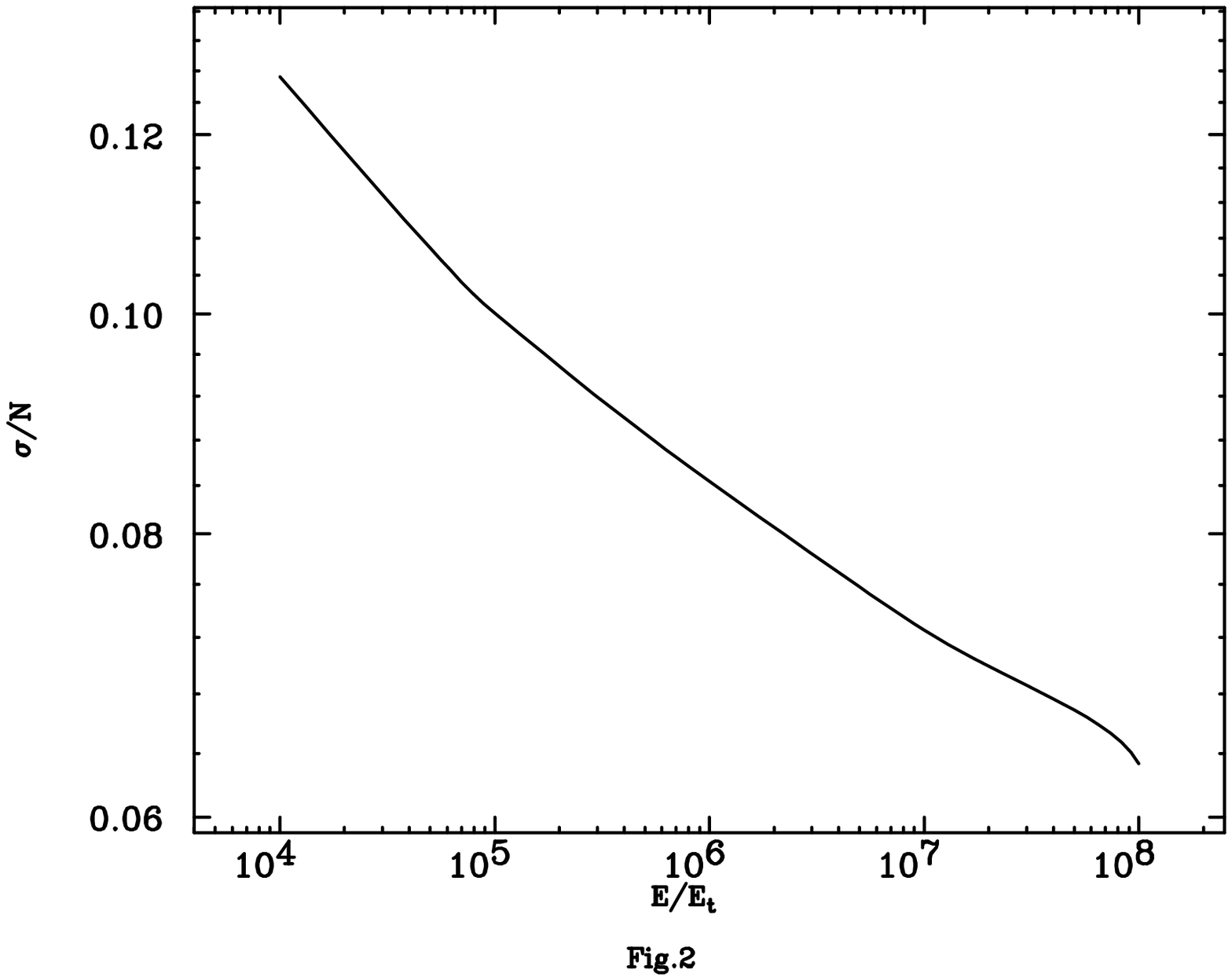,bbllx=60pt,bblly=200pt,bburx=600pt,%
bbury=600pt}}
\end{center}
\caption{
$\sigma / N $ evaluated at maximum depth as a function of the energy of the
primary particle}
\label{dos}
\end{figure}

As another example we calculated the value of $\sigma$ at maximum depth as a
function of the initial energy. We show it as the relative fluctuations, 
$\sigma/\langle N\rangle $, in figure (\ref{dos}). At low energy the relative
fluctuations are of the order of 15 \%, at higher energy it drops rapidly but
at a different slope. Note that the energy scale is logarithmic. This is also a
general feature.

\section{Conclusions} \label{final}

In this paper we have studied two different formalism to treat particle shower
fluctuations analytically. We have shown that both approaches lead to the same
solutions for the physical case of one particle in the initial state. Moreover,
the imposition of this equivalence between the two formalism simplifies the
possible forms of the solutions and yields directly an equation for the higher
moments in terms of the average number of particles. We have proven this in
generality for the academic case of zero dimensions and we have also proven it
to hold in the one dimensional case for the second order momentum. We have also
shown the total agreement between the solution of the analytical equations and
the results from a MonteCarlo simulation. Our results can be useful to
calculate properties of shower development at extremely high energy, where the
use of MonteCarlo programs is restricted by the huge number of particles
created in each shower.

It is straightforward to generalize the above results to the case of
electromagnetic cascades. For Furry's formalism, since the equations are 
linear, the step is trivial, although numerically more involved. The Janossy's
case was put forward by Janossy \cite{Janossy}. A more involved problem is the
hadronic cascades. We have to deal not only with processes of one particle
splitting into two, but also splitting into any number of other particles. This
additional problem complicates the counting of probabilities. Results for both
type of cascades are in preparation \cite{fluc_next}. 

\acknowledgments

I thank J.J. Blanco, M.C. Gonzalez-Garcia, F. Halzen, P. Lipari and E. Zas for
useful comments and discussions and for reading the manuscript. I also thank
M.C. Gonzalez-Garcia for her patience, insight and interest on the subject, in
the innumerable clarifying discussions which I infringed on her. I thank the
referee by pointing out references \cite{Kolmogorov,Kalmykov,Harris,Zatsepin}.
This work is supported by the EC, under contract number ERBCHBICT941658.


\end{document}